# Spin Peltier effect and its length scale in Pt/YIG system at high temperatures

Atsushi Takahagi[1], Takamasa Hirai[2], Ryo Iguchi[2], Keita Nakagawara[2,a)], Hosei Nagano[1], and Ken-ichi Uchida[2,3]*

[1]*Department of Mechanical Systems Engineering, Nagoya University, Nagoya 464-8601, Japan*
[2]*National Institute for Materials Science, Tsukuba 305-0047, Japan*
[3]*Institute for Materials Research, Tohoku University, Sendai 980-8577, Japan*
[a)]*Present address: Graduate School of Science, Tohoku University, Sendai 980-8578, Japan*
E-mail: UCHIDA.Kenichi@nims.go.jp

The temperature and yttrium-iron-garnet (YIG) thickness dependences of the spin Peltier effect (SPE) have been investigated using a Pt/YIG junction system at temperatures ranging from room temperature to the Curie temperature of YIG by the lock-in thermography method. By analyzing the YIG thickness dependence using an exponential decay model, the characteristic length of SPE in YIG is estimated to be 0.9 μm near room temperature and almost constant even near the Curie temperature. The high-temperature behavior of SPE is clearly different from that of the spin Seebeck effect, providing a clue for microscopically understanding the reciprocal relation between them.

In recent years, the field of spin caloritronics[1-3] has been rapidly developing by involving thermal effects in spintronics. In this field, various magneto-thermoelectric and thermo-spin effects unique to magnetic and spintronic materials have been discovered and investigated. Representative thermo-spin effects are the spin Seebeck effect (SSE)[4,5] and its reciprocal called the spin Peltier effect (SPE)[6,7] appearing in metal/magnetic-insulator junction systems. SSE (SPE) is the generation of a spin (heat) current from a heat (spin) current, where the spin current is carried by magnons in the magnetic insulator. In combination with the inverse (direct) spin Hall effect,[8,9] SSE (SPE) enables transverse heat-to-charge (charge-to-heat) current conversion in simple insulator-based systems. Because of their intriguing mechanism and technological advantages, SSE and SPE have attracted attention as next-generation energy harvesting and thermal management principles for spintronic devices, respectively.[10-12]

In order to clarify the physical mechanism of SSE and SPE, their temperature $T$, magnetic field $H$, and thickness dependences have been investigated systematically using the junction systems comprising a paramagnetic metal Pt and ferrimagnetic insulator yttrium iron garnet ($Y_3Fe_5O_{12}$:

YIG).[13-20] The $T$ dependence of SSE in Pt/YIG systems has been measured quantitatively in a wide temperature range from <10 K to above the Curie temperature of YIG[21] ($T_c$=550-560 K). In the high-temperature range, it has been reported that the inverse spin Hall voltage induced by SSE (SSE voltage) monotonically decreases with increasing $T$ and the $T$ dependence follows $\propto (T_c-T)^\alpha$ with $\alpha$ being the critical exponent.[21-23] The values of $\alpha$ obtained from various experiments were 1.5-3.0; the $T$ dependence of the SSE voltage for the Pt/YIG systems is convex downward (note that $\alpha$=3.0 was obtained when SSE was measured with applying a steady and uniform temperature gradient to the sample[21]). However, the origin of this dependence is yet to be clarified. On the other hand, SPE has been measured only below room temperature. Investigating the $T$ dependence of SPE at high temperatures quantitatively would help to clarify the origin of the thermo-spin conversion. Importantly, in 2020, Daimon *et al.* proposed a method to systematically measure the YIG thickness $t_{YIG}$ dependence of SPE using a Pt/wedged-YIG system.[24] By applying this method to the measurements in the high-temperature range, the $T$ dependence of the characteristic length of SPE can also be investigated.

In this paper, we report systematic investigations on the $T$ and $t_{YIG}$ dependences of SPE in the Pt/YIG junction system. We measured the spatial distribution of the temperature change induced by SPE in the Pt/wedged-YIG system by using the lock-in thermography (LIT) method[7,17,19,24] at various temperatures ranging from room temperature to around $T_c$. We obtained the $t_{YIG}$ dependence of SPE with high $t_{YIG}$ resolution at each temperature and estimated the characteristic length of SPE $l_{SPE}$ in the high-temperature range by analyzing the $t_{YIG}$ dependence. The obtained systematic dataset will help to elucidate the detailed mechanisms of SSE and SPE.

Figure 1(a) shows the schematic illustration of the sample used in this paper. The sample consists of a Pt strip with a thickness of 5 nm and width of 0.4 mm sputtered on an YIG/gadolinium gallium garnet ($Gd_3Ga_5O_{12}$: GGG) substrate with the YIG thickness gradient $\nabla t_{YIG}$. The substrate with $\nabla t_{YIG}$ was prepared by obliquely polishing a 25-μm-thick single-crystalline YIG (111) grown on a 0.5-mm-thick single-crystalline GGG (111) substrate by a liquid phase epitaxy method. The obtained $\nabla t_{YIG}$ was confirmed to be 6.4 μm/mm by a cross-sectional scanning electron microscopy [Fig. 1(c)]. In order to investigate the $T$ dependence, the sample was fixed on a stage with a heater and a temperature sensor by heat-resistant adhesive (Aron Ceramic D, Toagosei Co., Ltd.). Conducting wires were electrically connected to the ends of the Pt strip using heat-resistant conductive paste (MAX102, Nihon Handa Co., Ltd.). An insulating black ink was coated on the sample surface to increase the infrared emissivity and to make the emissivity uniform. To degas the sample and stage, they were heated up to 600 K for an hour in a high vacuum before the measurement. The electrical resistivity of the Pt strip was unchanged after heating and its $T$ dependence was similar to that obtained in a previous study.[25]

We measured the SPE-induced temperature change by using the LIT method [Fig. 1(d)]. A square-wave-modulated AC charge current with the amplitude of $J_0$=5 mA, frequency of $f$=5 Hz, and zero offset was applied to the Pt strip. To align the magnetization of YIG, an in-plane magnetic field **H** with the magnitude $\mu_0|H|$=100 mT was applied along the $x$ direction, where $\mu_0$ is the vacuum permeability. When the charge current is applied to the Pt strip, a spin current is generated by the spin Hall effect.[8,9] This spin current is injected into YIG at the Pt/YIG interface, causing the temperature change induced by SPE $\Delta T_{SPE}$ when the charge current direction is perpendicular to the magnetization of YIG.[6,7,17] As demonstrated in the previous studies, $\Delta T_{SPE}$ can be detected by the LIT method because $\Delta T_{SPE}$ changes its sign depending on the charge current direction.[7,17,19,24] The first-harmonic component of the temperature oscillation was extracted from thermal images taken with an infrared camera by Fourier analysis and converted into the lock-in amplitude $A$ and phase $\phi$ images. Since $\Delta T_{SPE}$ shows the $H$-odd dependence, we calculated $A_{odd}=|A(+H)e^{-i\phi(+H)}-A(-H)e^{-i\phi(-H)}|/2$ and $\phi_{odd}=-\arg[A(+H)e^{-i\phi(+H)}-A(-H)e^{-i\phi(-H)}]$ from the thermal images measured with applying positive and negative magnetic fields. The SPE-induced temperature change was obtained by $\Delta T_{SPE}=A_{odd}\cos\phi_{odd}$ since the time delay due to thermal diffusion is negligibly small in the LIT-based SPE measurements.[7,17,19] Figure 1(b) shows a steady-state infrared image of the black-ink-coated sample at 314 K, where the YIG film with $\nabla t_{YIG}$ exists below the white dotted line. In the region above the line, no SPE signal should be generated above room temperature because of the absence of YIG. The area surrounded by an orange dotted rectangle shows the position and size of the Pt strip. Based on the $\nabla t_{YIG}$ value, the $t_{YIG}$ resolution in our setup was obtained to be 30 nm/pixel. During the LIT measurements, the temperature of the sample surface was monitored with the infrared camera in a high vacuum of $<4\times10^{-4}$ Pa through a $CaF_2$ window with high infrared transparency. The small reduction of the infrared emission intensity from the sample due to the $CaF_2$ window was corrected based on a calibration curve, which was measured by using a resistance temperature sensor as a dummy sample and by comparing the sensor and thermal image values.

Figures 2(a) and 2(b) show the $A_{odd}$ and $\phi_{odd}$ images for the Pt/wedged-YIG sample at $T$=314 K. The current-induced temperature change appears only in the region with the Pt strip on the YIG film. Since $A_{odd}$ and $\phi_{odd}$ are the $H$-odd components, the contribution from the field-independent Peltier effect is eliminated. The magnitude ($A_{odd}$) and sign ($\phi_{odd}$) of the observed temperature modulation is consistent with the SPE signal reported previously in the Pt/YIG systems.[24] No temperature modulation is generated in the region without YIG, confirming that the ordinary

Ettingshausen effect in Pt is negligibly small. These features indicate that this temperature change is due to SPE. Figures 2(c) and 2(d) show the profile of the $A_{\text{odd}}$ and $\phi_{\text{odd}}$ signals in the $y$ direction. In the region with finite $t_{\text{YIG}}$, the SPE signal monotonically increases with increasing $t_{\text{YIG}}$ and saturates when $t_{\text{YIG}}>5$ μm. The spatial distribution of the SPE signal is also consistent with the previous result.[24]

Figures 3(a) and 3(b) show the $A_{\text{odd}}$ and $\phi_{\text{odd}}$ images for various values of $T$ in the high-temperature range. The images show that the SPE signal disappears at 552 K, around $T_c$ of YIG [note that the electrical resistivity of the Pt strip exhibits no anomaly around $T_c$, as shown in the inset to Fig. 3(c)]. The $T$ dependence of $\Delta T_{\text{SPE}}$ at each $t_{\text{YIG}}$ was calculated by averaging the $A_{\text{odd}}$ and $\phi_{\text{odd}}$ data along the $x$ direction over a length of 0.4 mm. At $t_{\text{YIG}}=5$ μm, where the SPE signal reaches the saturation value, the magnitude of the SPE signal is almost constant up to 400 K but monotonically decreases with increasing $T$ for $T>400$ K [Fig. 3(c)]. The similar $T$ dependence of $\Delta T_{\text{SPE}}$ was obtained also in the small $t_{\text{YIG}}$ region in which the SPE signal does not saturate.

To investigate the $T$ dependence of the characteristic length of SPE $l_{\text{SPE}}$ in YIG, we analyzed the $t_{\text{YIG}}$ dependence of the SPE signal at each temperature. Here, we adopt a phenomenological exponential decay model as a simplest analysis:

$$\Delta T_{\text{SPE}} \propto 1 - \exp\left(-\frac{t_{\text{YIG}}}{l_{\text{SPE}}}\right). \tag{1}$$

Figure 4(a) shows the experimental and fitting results for various values of $T$. When the SPE signal is sufficiently large, the experimental results are well fitted by Eq. (1) in the whole thickness range, where the coefficient of determination is $R^2>0.8$ for $T<520$ K. However, for $T>520$ K, the fitting accuracy is poor ($R^2<0.8$) in the small $t_{\text{YIG}}$ region because of the small magnitude of the SPE signal. Figure 4(b) shows the $T$ dependence of $l_{\text{SPE}}$ and $R^2$. The $l_{\text{SPE}}$ value near room temperature (314 K) is estimated to be 0.9 μm, which is comparable to the values obtained in the previous studies on SPE.[24,26,27] We found that $l_{\text{SPE}}$ remains almost constant as the temperature increases.

Next, we compare the $T$ dependence of the SPE signal with that of the SSE voltage. If the reciprocal relation between SPE and SSE and $T$-independent $l_{\text{SPE}}$ are assumed, the SSE voltage normalized by the applied temperature difference can be compared with the factor proportional to $(\kappa/T)(\Delta T_{\text{SPE}}/j_c)$, where $\kappa$ is the thermal conductivity of YIG.[20] Figure 4(c) shows the $(\kappa/T)(\Delta T_{\text{SPE}}/j_c)$ values as a function of $T$ at $t_{\text{YIG}}=5$ μm, where the $T$ dependence of $\kappa$ is estimated from the data in Ref. 21. The magnitude of $(\kappa/T)(\Delta T_{\text{SPE}}/j_c)$ almost linearly decreases with increasing $T$, which is different from the behavior of the SSE voltage.[21] In order to quantify the $T$ dependence of SPE, $(\kappa/T)(\Delta T_{\text{SPE}}/j_c)$ is fitted by

$$\frac{\kappa}{T}\frac{\Delta T_{\text{SPE}}}{j_{\text{c}}} \propto (T_{\text{c}} - T)^{\beta}. \tag{2}$$

The ($\kappa/T$)($\Delta T_{\text{SPE}}/j_{\text{c}}$) data is well fitted with the critical exponent of $\beta$=1.1 [Fig. 4(c)]. We observed the same $T$ dependence of the SPE signal with $\beta$=1.1 also in a Pt-film/YIG-slab junction system with the single-crystalline YIG slab grown by a flux method. This result is clearly different from $\alpha$=1.5-3.0 for SSE in the high-temperature range.

Finally, we discuss the possible reason of the different $T$ dependence of SPE and SSE. We emphasize again that the same $T$ dependence of the SPE signal was obtained not only in the YIG film grown by the liquid phase epitaxy method but also the YIG slab grown by the flux method, indicating that the difference in the growth method of YIG is irrelevant to the different behaviors between SPE and SSE. Now recall the fact that the thermo-spin conversion by SPE and SSE has the magnon frequency dependence.[13] In the case of SSE, a temperature gradient applied to the Pt/YIG system induces magnon spin currents and low-frequency subthermal magnons dominantly contribute to the SSE voltage.[13,14,28] On the other hand, in the case of SPE, the spin Hall effect in Pt induces magnon spin currents in YIG and the magnon frequency dominantly contributing to the SPE-induced temperature change is unclear. If the magnon frequency excited by the spin Hall effect in SPE is different from that excited by the temperature gradient in SSE, the $T$ dependence and characteristic length of these phenomena can be different from each other. In fact, the characteristic length for SPE in Fig. 4(b) is comparable to that estimated in the previous SPE experiments[24,26,27] but smaller than that estimated in the previous SSE experiments.[15,29-33] This situation indicates that the Onsager reciprocal relation between SPE and SSE is not simple and its magnon frequency dependence should be taken into account. To clarify the spectral nature of SPE, in addition to the $T$ and $t_{\text{YIG}}$ dependences, the high-magnetic-field response of SPE and SSE should be compared systematically in a wide temperature range, which is one of the remaining tasks in the study of these phenomena.

In summary, we investigated the $T$ and $t_{\text{YIG}}$ dependences of SPE by the LIT method. The SPE signal in the Pt/YIG system was found to monotonically decrease with increasing $T$ when $T$>400 K and disappear around the Curie temperature of YIG. By fitting the observed $t_{\text{YIG}}$ dependence of the SPE signal by the exponential decay model, the characteristic length of SPE was estimated to be 0.9 μm near room temperature (314K) and be almost constant as the temperature increases. The SPE-related factor that can be compared with the SSE voltage, ($\kappa/T$)($\Delta T_{\text{SPE}}/j_{\text{c}}$), shows the ($T_{\text{c}}$–$T$)$^{1.1}$ dependence, which is significantly different from the $T$ dependence of the SSE voltage in the Pt/YIG system: ($T_{\text{c}}$–$T$)$^{\alpha}$ with $\alpha$=1.5-3.0. The results reported here highlight the difference in the thermo-spin conversion between SPE and SSE and suggest the magnon-frequency-dependent nature in the reciprocal relation between them.

**Acknowledgments**

The authors thank M. Isomura for technical supports. This work was supported by CREST "Creation of Innovative Core Technologies for Nano-enabled Thermal Management" (No. JPMJCR17I1) from JST, Japan; Grant-in-Aid for Scientific Research (B) (No. 19H02585) and Grant-in-Aid for Scientific Research (S) (No. 18H05246) from JSPS KAKENHI, Japan; NEC Corporation; and NIMS Joint Research Hub Program.

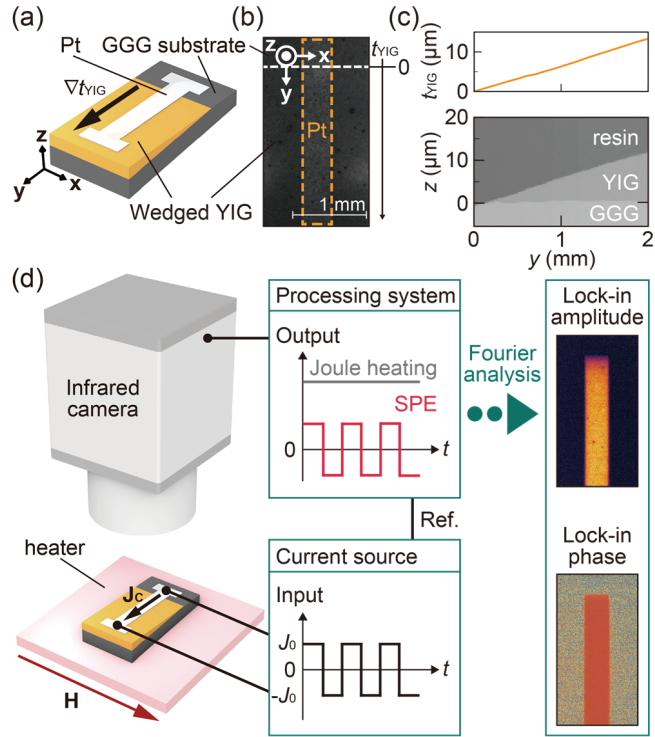

**Fig. 1.** (a) Schematic of the Pt/wedged-YIG system used for the SPE measurement. The YIG film has a thickness gradient $\nabla t_{YIG}$ along the $y$ direction. (b) Steady-state infrared image of the Pt/wedged-YIG system. The area below the white dotted line corresponds to the area with the finite YIG thickness $t_{YIG}$. The area surrounded by the orange dotted rectangle corresponds to the area with the Pt film. (c) $t_{YIG}$ profile and cross-sectional image of the wedged-YIG/GGG substrate without the Pt layer, measured by scanning electron microscopy. The samples used for the SPE and scanning electron microscopy measurements were cut from the same wafer. (d) Schematic of the LIT method for measuring SPE. The base temperature $T$ of the sample was controlled by a heater attached to the stage and monitored with an infrared camera.

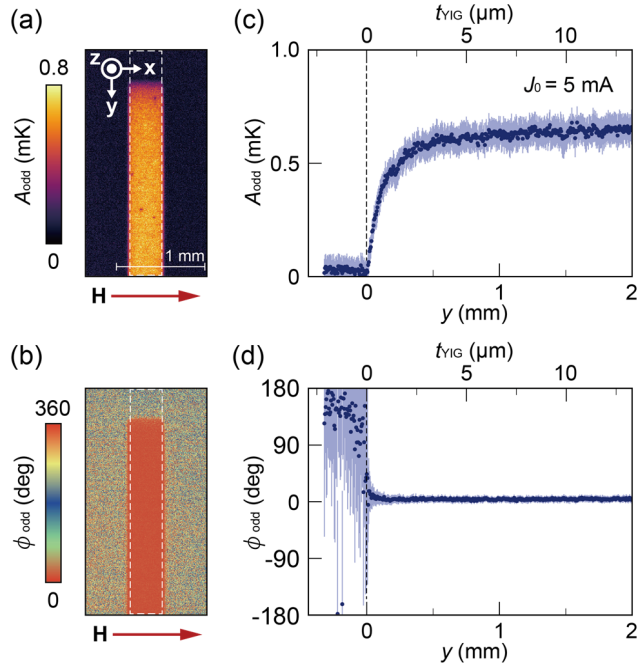

**Fig. 2.** (a),(b) Lock-in amplitude $A_{\mathrm{odd}}$ and phase $\phi_{\mathrm{odd}}$ images for the Pt/wedged-YIG system at $T=314$ K, magnetic field of $\mu_0|H|=100$ mT, and square-wave charge current amplitude of $J_0=5$ mA. $\mu_0$ is the vacuum permeability. (c),(d) $y$-directional profile and corresponding $t_{\mathrm{YIG}}$ dependence of the $A_{\mathrm{odd}}$ and $\phi_{\mathrm{odd}}$ signals for the Pt/wedged-YIG system at $T=314$ K. The $A_{\mathrm{odd}}$ and $\phi_{\mathrm{odd}}$ profiles are obtained by averaging the raw data in the area defined by the white dotted squares in (a) and (b), respectively.

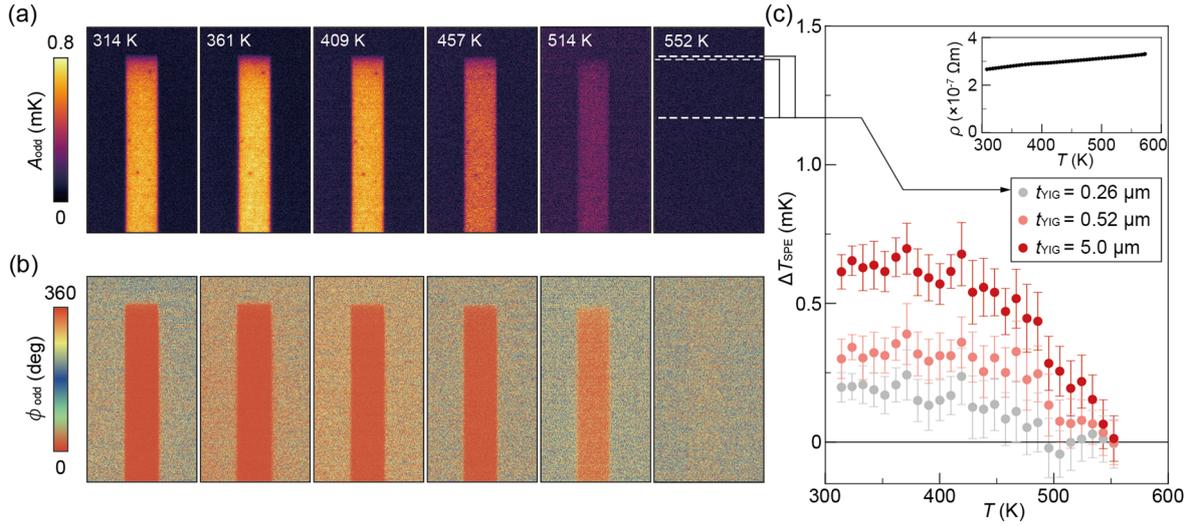

**Fig. 3.** (a),(b) $A_{odd}$ and $\phi_{odd}$ images for the Pt/wedged-YIG system for various values of $T$ at $\mu_0|H|$=100 mT and $J_0$=5 mA. (c) $T$ dependence of $\Delta T_{SPE}=A_{odd}\cos\phi_{odd}$ for various values of $t_{YIG}$. The data points are obtained by averaging the raw data along the $x$ direction over 0.4 mm on the positions of the white dotted lines depicted in the rightmost $A_{odd}$ image in (a). The error bars represent the standard deviation of the data. The inset to (c) shows the $T$ dependence of the resistivity $\rho$ of the Pt film. The $\rho$ values were measured using a 5-nm-thick Pt film sputtered on a single-crystalline YIG without $\nabla t_{YIG}$ by the four-terminal method.

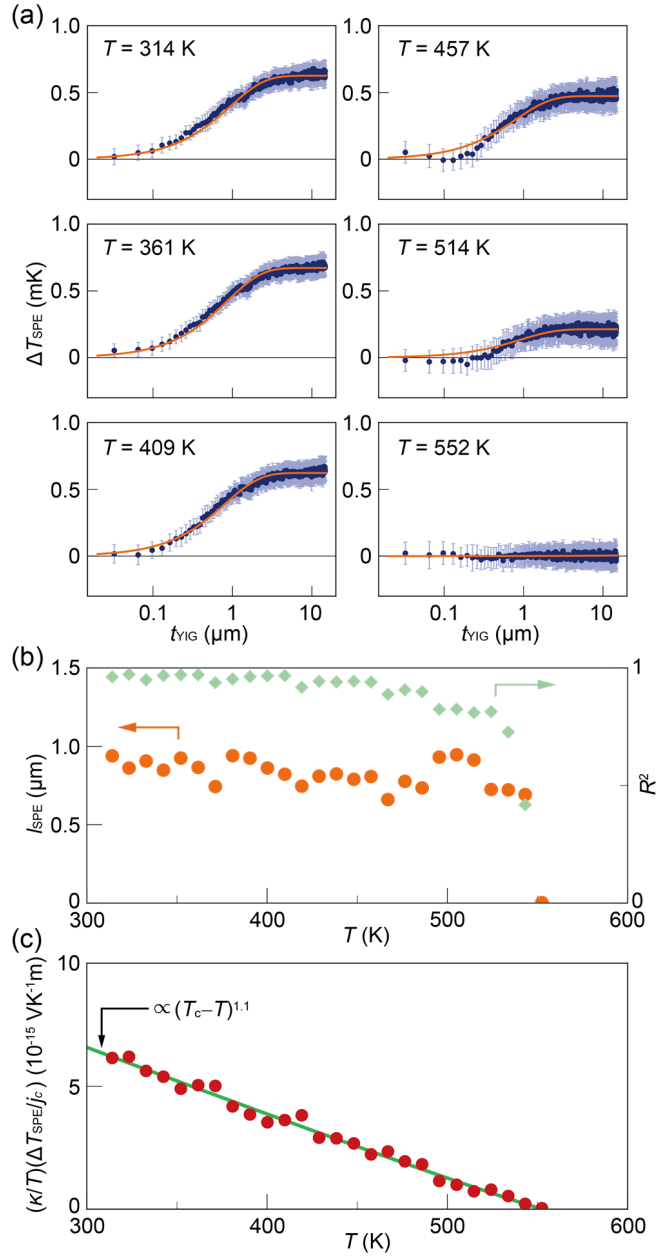

**Fig. 4.** (a) $t_{YIG}$ dependence of $\Delta T_{SPE}$ (blue) and fitting curves (orange) for various values of $T$ at $\mu_0|H|$=100 mT and $J_0$=5 mA. (b) $T$ dependence of the characteristic length of SPE $l_{SPE}$ (orange) and coefficient of determination of the fitting $R^2$ (green), obtained by fitting the $t_{YIG}$ dependence of $\Delta T_{SPE}$ at each temperature with Eq. (1). (c) $T$ dependence of $(\kappa/T)(\Delta T_{SPE}/j_c)$, where $\kappa$ and $j_c$ are the thermal conductivity of YIG and charge current density, respectively. The $T$ dependence of $\kappa$ is estimated from the data in Ref. 21. The green line shows the fitting results using Eq. (2).